# Separation Effect of Early Visual Cortex V1 Under Different Crowding Conditions：A TMS Study

Xieyi Liu, Junjun Zhang, Ling Li*

Key Laboratory for Neuroinformation of Ministry of Education, School of Life Science and Technology, University of Electronic Science and Technology of China, Chengdu 610054, P. R. China

*Corresponding author: liling@uestc.edu.cn

**Abstract:** The visual crowding makes it difficult to identify the patterns in peripheral vision, but the neural mechanism for this phenomenon is still unclear because of different opinions. In order to study the separation effect of V1 under different crowding conditions, single-pulse transcranial magnetic stimulation is applied within the right V1. The experimental design includes two factors: TMS intensity (10%, 65%, and 90% of the phosphene threshold) and crowding (high and low) conditions. The accuracy results show that there is a strong interaction between crowding condition and TMS condition. When the TMS stimulation intensity is lower than the phosphene threshold, more crowding will be perceived under the high crowding condition, and less crowding will be perceived under the low crowding condition. The above results conclude that the high and low crowding condition separate by TMS stimulation. The results support the assumption that the crowding is related to V1 and occurs in the visual coding phase.

## 1  Introduction

When a target is presented with nearby flankers in the peripheral visual field, it becomes harder to identify, which is referred to as crowding [1]. Crowding is an important obstacle to object recognition. For ordinary people, peripheral vision is more susceptible to crowding, while central vision is basically unaffected [2,3]. Studying the neural mechanism of crowding and related brain regions is helpful to find a breakthrough to solve diseases such as dyslexia which is related to visual crowding.

At present, the reasons for crowding effect can be classified into two categories in general [4]: (1) It is considered that the crowding effect occurs in the visual coding phase. According to this view, extraction or integration of target stimuli has been suppressed or interfered during the visual coding phase [12,14,22,11]; (2) The target stimuli was not affected during the visual coding phase, and the participants perform not quite well because of insufficient attention resolution [8,24-26].

The neural mechanisms involved in the visual crowding are still uncertain, and some studies supporting that visual crowding effect in the visual coding phase is associated with V1. Both Blake et al. [12] and Ho et al. [14] found that the crowding decreases the adaptational aftereffect of the target grating, which indicates that the crowding occurs before the orientation adaptation, so that they infer the lower visual cortex V1 is the position of crowding. Millin et al. [15] used fMRI to explore the nerve area related to crowding. In the state without crowding, when the target and flankers appear at the same time, the level of blood oxygen (blood oxygen level dependent, BOLD) signal is stronger than the contrary situation. However due to the suppression between the target and flankers, the growth of signal reduces when target and flankers appear at the same time instead of only flankers appear in crowding conditions. They found that this inhibitory effect can be produced at V1, and the stronger crowding of the stimulus, the stronger suppression of V1. Chen et al [18]. used ERPs and fMRI to study the neural mechanism of crowding from two perspectives: time and space. The signal difference between C1 and V1 was related to the behavioral crowding of participants. The difference between C1 suppression and V1 signal are stronger when behavioral crowding is stronger. This shows that crowding occurs in V1 and is regulated by attention. Although the studies above all support that V1 is the region where crowding occurs, the fMRI studies of Arman et al. [16] and Bi et al. [17] did not find significant crowding effects at V1, instead they found the lowest cortical area associated with crowding is V2.

At present, the visual crowding is carried out by fMRI and ERP experiments. Those two experimental methods can only ensure that the changes in signal in V1 occur simultaneously with the stimulus, but the exact correlation cannot be proved. In order to further study the correlation between crowding and V1, we performed a transcranial magnetic stimulation (TMS) experiment. TMS is a technique that produces focal, transient and fully reversible disruptions in brain activity by delivering strong magnetic pulses to the cortex that pass through the skull and depolarize the underlying neurons of particular areas in the brain [13, 10]. This method has been proved to be a useful research approach to assess the functional and causal role of specific cortical areas in cognition [19]. TMS experiment can be used to observe changes in results under different crowding conditions by giving short-term pulse stimulation to specific brain regions. Abrahamyan et al. [20] found that TMS-induced neuronal activity can be combined with stimulus-induced activity to enhance visual perception when stimulating the V1 region with intensity below the phosphene threshold according to TMS experiments. Based on the above conclusion, we plan to stimulate V1 with different strengths of TMS to study its effects on high and low crowding conditions. We hypothesized that stimulating the V1 region with a single-pulse TMS intensity below the phosphene threshold can augment visual perception. Besides, the visual crowding occurs in the visual coding stage. Under the crowding condition, more crowding will be felt when the visual perception increases. So that it is more difficult to integrate and extract the features of stimulate during the visual coding phase, resulting in the increase of crowding.

In previous studies, it has been found that the appearance of phosphene is related to V1 and V2. The precise Montreal Neurological Institute (MNI) coordinate is found by the navigation system of the coil and considered to be the accurate position of V1. Then, the crowding experiment is started. This method can solve the difficulty of locating V1.

## 2 Materials and methods

### 2.1 Participants

In the experiment, we tested a total of 18 participants (7 male). All participants are right-handed and reported normal or corrected-to-normal vision. Ages range from 19 to 25 years. All participants conform to the following selection requirements: 1) no pregnancy; 2) no metal implants in the body; 3) no history of mental illness. The study was approved by the UESTC Ethics Board. Written informed consent was obtained from each participant prior to being tested.

## 2.2 Stimuli

All the targets and flankers are circular sinusoidal gratings (diameter: 2.12°; mean luminance: 61.47cd/m$^2$). The background luminance is 61.47cd/m$^2$ [18]. Previous studies reported that the crowding in the upper field of view is stronger than in the lower field of view. Therefore, the stimulus was present in the upper left field of view and centered at 8° eccentricity in the upper left visual quadrant. The orientation of the target was $45° \pm \theta$, either left or right rotated. The orientations of the flankers were independently and randomly selected from 0° to 180° for each trial. The targets and flankers' settings include high crowding (center distance is 2.33°) and low crowding (center distance is 4.53°).

## 2.3 Apparatus

TMS pulse stimulation is supported by using Magstim super rapid magnetic stimulator and air-cooled figure-of-eight coil (Magstim Company Limited, Whiteland, UK). The coil manually holds the handle horizontally to the right side of the participant and tangent to the scalp by using the online visualization function of the BrainSight frameless stereotaxy system (BrainSight Frameless, Rogue Research, Montreal, QC, Canada).

The stimulation position's MNI coordinates were targeted via the BrainSight stereotaxic neuronavigator (Rogue Research, Montreal, QC, Canada), equipped with a Polaris Vicra position sensor system. With the help of Psychtoolbox [5,6], the experimental program was written by MATLAB (MathWorks) and connected to the TMS instrument to realize simultaneous triggering of TMS pulses and stimuli. The visual stimulus is displayed on the computer monitor (refresh rate: 60Hz; resolution: 1024×768) and the background is gray.

Three TMS stimulation intensities below the phosphenes threshold were set in the experiment：10%, 65% and 90% of the phosphenes threshold respectively, and were recorded as 10% TMS, 65% TMS and 90% TMS. The single-pulse TMS stimulation coincided with the target grating. The three TMS conditions and the sequence of experiment 1 and experiment 2 were randomly among participants. Because the position of visual cortex V1 is deep and has great difference among people, it is difficult to locate it according to the precise coordinates. The Polaris Vicra position sensor system is used to find the coordinates of that point, which is considered to be an accurate position of V1, which solves the difficulty of V1 positioning.

# 3  Design

## 3.1  Phosphenes threshold test

The position of the coil was fixed according to the MNI coordinates obtained before with the help of transcranial magnetic stimulation localization cap, and then the stimulation intensity of TMS was reduced by 2%-3% step size until participants ensure that phosphenes phenomenon disappeared. The minimum stimulus intensity that produces the phosphenes reaction was record as phosphenes threshold.

## 3.2  Orientation discrimination threshold test

$\theta$ was the orientation discrimination threshold (85% correct) for the target in the experiments 1 and 2. To measure the discrimination threshold, a stimulus (Low or High) was presented for 250ms. The orientation of the target was either $45°+\theta$ or $45°-\theta$. Participants were asked to judge the orientation of the target relative to 45° (clockwise or counterclockwise) and press the left or right button of the keyboard to report the judgment result. The $\theta$ varied trial by trial and was controlled by the QUEST staircase [29]. We use the same orientation discrimination threshold, which is calculated by the average value of orientation discrimination threshold in both conditions (Low and High).

## 3.3  Experiment Procedures

Before we start Section 1 and Section 2, we test all participants to check whether they can elicit phosphenes while stimulating the right occipital lobe. The participant sat in a dimly lit room and dark adapted for 5 min and they were asked to relax and look at the green dot which diameter is 0.5 cm in the center of the black screen. The participant's head was supported by a sitting posture corrector and the viewing distance is 62 cm from the monitor. The TMS coil was positioned with the handle oriented to the right side of the participant in a horizontal direction and is tangent to the scalp. The coil was initially placed against the back of the participant's head, with the center over an area 3 cm above the inion and 2 cm right. Single pulses were delivered with intensities reaching 80%-85% of the stimulator's output and the initial stimulus intensity is determined by the maximum stimulus intensity that the participant can withstand without causing uncomfortable reactions. After being stimulated five times in five seconds at the same position, the participant's will be asked whether they think that there is a change surrounded the green dot or around the screen, including the appearance of white flash, white dots, white lines, etc. The coil was moved in steps of 0.5–1 cm and the accurate position will be found which can induce five times phosphenes reaction [20]. Salminen-Vaparanta et al. [21] found that V1 and V2 have similar abilities to produce conscious perception that the phosphene induced by stimulating V1 is brighter than stimulating V2 through TMS and fMRI studies. Therefore, in present paper, we first find the position of the participants which can produce phosphene in the right occipital cortex before the experiment, and compared with the surrounding position to find point that can induce the brightest phosphene phenomenon. This accurate position is compared with the position nearby, in order to find the point that can cause the most obvious and brightest phosphenes phenomenon. The navigator and the fMRI structure image of the participant are used to get the MNI coordinates of the phosphenes point and recorded it as TMS stimulation coordinates for subsequent experiments. The coordinates of the phosphene point were obtained by the navigator and the fMRI structure of the participants, and recorded as the TMS stimulation site of the experiment.

In the Section 1, the target between flankers are high crowding and the target between flankers are low crowding in the Section 2. The process of Section 1 and Section 2 are the same (see Figure 1). Before each trial, there will be a blank screen with time of 750-1250ms randomly, and then the stimulus will appear 250ms and disappear after. During the whole process of the experiment, the participant is required to fix their eyes on the black dot at the center of the screen. There will be a fixed white dashed circle to indicate the position of the target stimulus. The participant required to pay attention to the white dashed circle, and judge the direction of the target when it appears. Participants need to judge the direction of the target stimulus relative to 45° (clockwise or counterclockwise). The rotation angle $\theta$ for each participant is determined by the orientation discrimination threshold measured before the experiment. Participants need to press the left or right button of the keyboard to report the result. After the participant responded, the blank screen before the next trial will display. During the experiments, the program will record the button status, reaction times and accuracy of the participant.

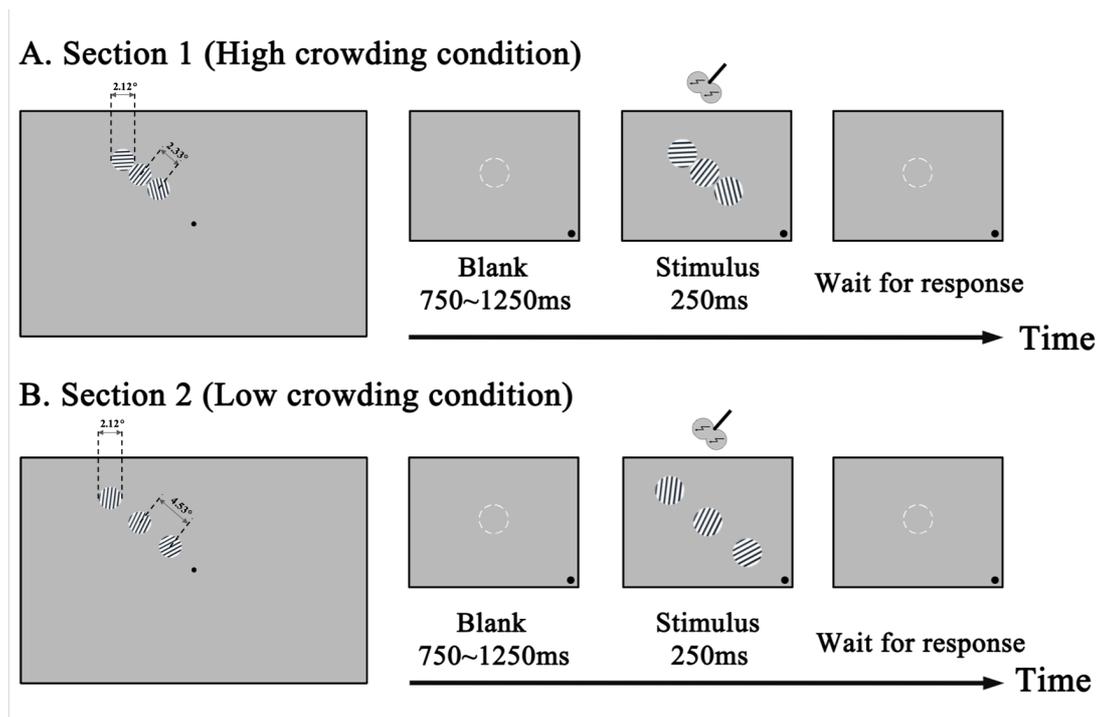

Figure 1. Experiment design. The figure on the left shows the screen of each section, and the figure on the right shows the stimuli are presented in the upper left visual quadrant and the protocol of each section. In two sections（Section 1: High crowding condition；Section 2: Low crowding condition）, participants are tested under three TMS condition. Participants need to fixate the black dot in the center of the monitor and pay attention to upper left field of vision. The grating stimulus is presented for 250ms in upper left field of vision. During the experiment, a single pulse of TMS is delivered to V1 at the same time as the stimulus appears.

## 4  Results

All behavioral results were normally distributed (Kolmogorov–Smirnov test values>0.05). Repeated-measures analyses of variance (ANOVAs) is applied to reaction times (RTs) and accuracy (ACC) results (Bonferroni's corrected). Mauchly's sphericity test was used for compound symmetry. When the compound symmetry was not satisfied, Greenhouse–Geisser correction and post hoc t-test are used. In all tests, an α threshold of 0.05 for assessing statistical significance is consistent.

The average phosphene threshold of 18 participants in the experiment is 73% and the average orientation discrimination threshold is 6.5125°. The average MNI coordinate of phosphene position of 18 participants is founded: x=29.9±9.13mm; y=-102.85±8.18 mm; z=17.29±19.30 mm.

### 4.1  RTs Results

The 2×3 repeated ANOVA reveal only a main effect of crowding condition [$F_{(1,17)}$=21.68, $p<0.001$] (see Figure 2). There is no main effect of TMS condition [$F_{(2,16)}$=0.072, $p$=0.931]. There has no significant interaction effect between crowding condition and TMS condition [$F_{(2,16)}$=0.291, $p$=0.751]. Under the same crowding condition, post hoc t-test is used in three TMS conditions (10% TMS, 65% TMS, 90% TMS). There is significant difference between high and low crowding condition with each TMS condition [10% TMS, $t_{(17)}$=2.504, $p$=0.023; 65% TMS, $t_{(17)}$=2.346, $p$=0.031; 90% TMS, $t_{(17)}$=3.097, $p$=0.007].

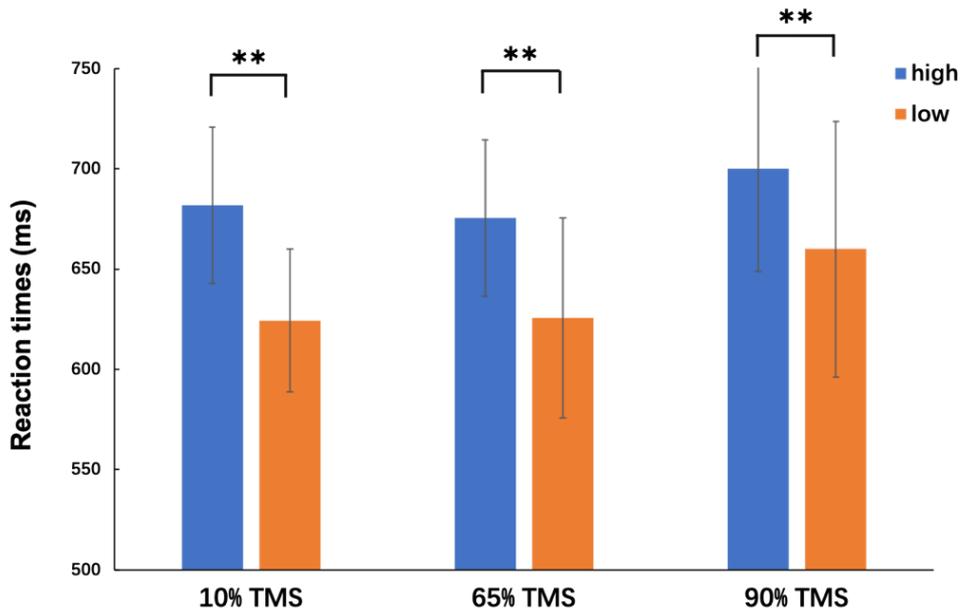

Figure 2. Mean reaction times (RTs) are shown for three TMS conditions with low and high conditions. Asterisks mark significant post hoc t-test ($p < 0.05$)

## 4.2 Accuracy Result

From the average accuracy of the three TMS conditions, the average accuracy of the high condition (81.67%) is greater than the average accuracy of the low condition (78.83%) with 10% TMS condition, and the average accuracy of the high condition is less than the average accuracy of the low condition with 65% TMS condition (77.39% < 82.28%) and 90% TMS condition (77.5% < 83.11%). The 2×3 ANOVA reveal only a main effect of crowding condition [$F_{(1,17)}$ = 4.952, $p<0.05$]. There is no main effect of TMS condition [$F_{(2,16)}$=0.100, $p$=0.906]. There has significant interaction effect between crowding and TMS condition [$F_{(2,16)}$=18.393, $p<0.001$].

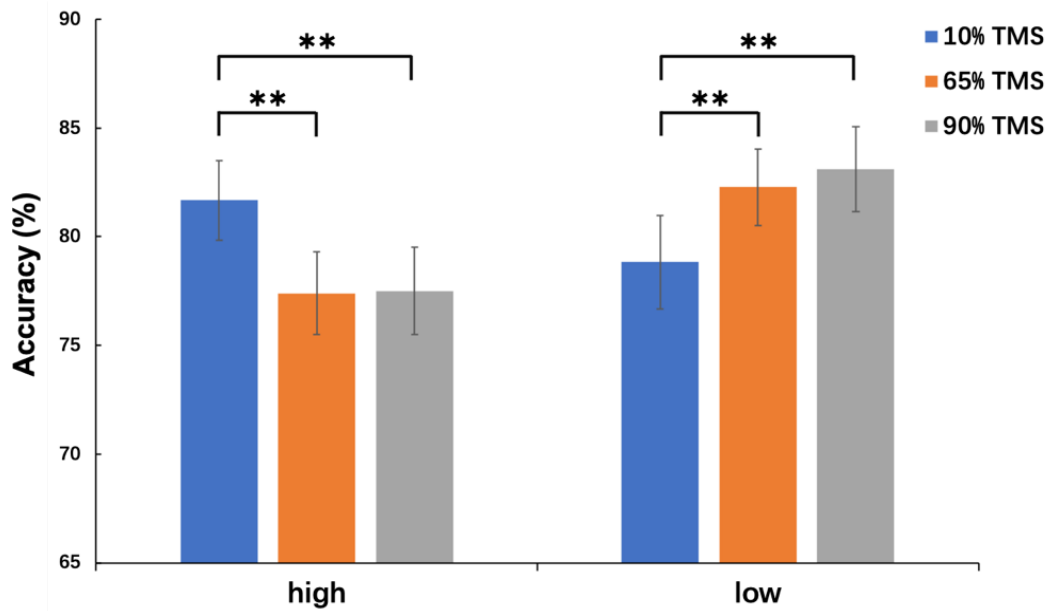

Figure 3. Mean accuracy are shown for two crowding conditions with three TMS conditions (10% TMS, 65% TMS, 90% TMS). Asterisks mark significant post hoc t-test (p < 0.05)

Under the same TMS condition, post hoc t-test is used in two crowding conditions (high, low). The result shows that there has a significant difference in accuracy of two crowding condition (high, low) with 10% TMS condition [$t_{(17)}$=2.136, p<0.05], 65% TMS condition [$t_{(17)}$=-3.076, p<0.005], 90% TMS condition [$t_{(17)}$=-3.076, p<0.01].

Under the high crowding condition, Bonferroni corrected is used in three TMS condition (10% TMS, 65% TMS, 90% TMS) (see Figure 3). There has a significant performance difference between 10% TMS and 65% TMS [$t_{(17)}$=2.954, p=0.027], 10% TMS and 90% TMS [$t_{(17)}$=3.828, p=0.003]. However, there are no difference between 65% TMS and 90% TMS [$t_{(17)}$=-0.079, p=2.814]. Similarly, there has a significant performance difference between 10% TMS and 65% TMS [$t_{(17)}$=-2.699, p=0.045], 10% TMS and 90% TMS [$t_{(17)}$=-3.341, p=0.012] and no difference between 65% TMS and 90% TMS [$t_{(17)}$=-0.531, p=1.806] with low crowding condition. Conclusion is obtained that high TMS stimulus intensity impact is more obvious then low TMS stimulus intensity on crowding.

## 5 Discussion

The ANOVA analysis for the reaction times shows that in the case of different crowding conditions, regardless of the TMS condition, the RTs in the high crowding condition is significantly smaller than the RTs in the low crowding condition. The crowding condition has a significant main effect, indicating that crowding phenomenon exist under different TMS conditions with different intensity. The 2×3 repeated measures ANOVA analysis of accuracy has a significant main effect of the crowding condition, indicating that from the situation of three TMS conditions, the overall accuracy in the high condition is less than the accuracy in the low condition. There is a strong interaction between crowding condition and TMS condition, means that stimulation on V1 with TMS inhibited the high condition and promoted the low condition for participants to recognize the target. Those two effects are significant.

Many previous studies have suggested that the crowding is related to the early visual cortex and is likely to be related to V1, but these results are previously confirmed by fMRI experiments [15,18]. The use of gratings to simulate crowding, combining with the TMS pulse stimulation online, will have an impact on the V1 zone. Previous studies have proved that the visual sensitivity will improve when stimulating V1 with intensity below the phosphenes threshold [20]. The crowding effect does not affect the participants' detection of the target, only affects the recognition of the target. Therefore, the increase of visual sensitivity will feel more crowding in the condition of high crowding, and the identification of the target is more difficult. Similarly, less crowding will be felt in the condition of low crowding and improve the ability to recognize the target, which is accord with our experimental results [9, 23].

One explanation for the crowding of previous studies is based on long-range horizontal connections between different populations of neurons, which may be promotion or suppression [27,28]. Stettler et al. showed that the horizontal connections cover portions of V1 representing regions of visual space up to eight times larger than classical receptive fields [7]. In our experiments, the spatial extent of the 8° eccentricity of neurons can completely cover two gratings. This long-range horizontal connections links neurons with similar response characteristics, including neurons with similar spatial frequencies in the experiment [27], which demonstrates that the crowding in the experiment is certain because of the intrinsic structure of the peripheral visual system. In the case of high crowding, the establishment of long-range horizontal connections may become easier. TMS stimulation below the threshold of phosphenes will establish more long-range horizontal connections, affecting feature extraction, resulting in more crowding.

Conversely, in the case of low crowding, long-range horizontal connections are less stable, and TMS may break these connections, improve feature extraction, and reduce crowding.

Millin et al. [15] believe that that the main reason for the occurrence of crowding is the lack of feature integration and segmentation in peripheral vision at the earliest stage of cortical processing, mainly in V1. The visual system's perception of graphical features is to integrate certain features in the graphic, and to prevent these features from being combined with other features. This visual extraction and integration of graphical features occurs in the visual coding phase. Our experimental results support the conclusion that visual crowding occurs during the visual coding phase. From the perspective of visual sensibility and visual feature extraction in the visual coding phase, the improvement of visual sensitivity will make it easier to integrate the features of the target extracted from the peripheral visual field with the features of flankers in high crowding condition, which is difficult to segment. Under the low crowding condition, it is easier to extract features of the target and the crowding is reduced.

Our experimental results did not rule out the possibility of crowding occurring in multiple visual processing stages. Because this phenomenon is a complicated process, it has relationship with the stimulation and experimental methods used in the experiment. Our work demonstrates the relationship between crowding and V1 and the separation effect of TMS from different crowding conditions, and is benefit to further clarify the neural mechanism of visual crowding effect.

## Acknowledgement

This work was supported by the National Natural Science Foundation of China under grants 61773096, 61673087 and 61773092.